# DOES A NON-MAGNETIC SOLAR CHROMOSPHERE EXIST?

Mats Carlsson

Institute of Theoretical Astrophysics, P.O. Box 1029, Blindern, N–0315 Oslo, Norway

AND

Robert F. Stein

Department of Physics & Astronomy, Michigan State University, East Lansing, MI 48824, USA

## ABSTRACT

Enhanced chromospheric emission which corresponds to an outwardly increasing semiempirical temperature structure can be produced by wave motion without any increase in the mean gas temperture. Hence, the sun may not have a classical chromosphere in magnetic field free internetwork regions. Other significant differences between the properties of dynamic and static atmospheres should be considered when analyzing chromospheric observations.

*Subject headings:* hydrodynamics – radiative transfer – shock waves – Sun:chromosphere

## 1. INTRODUCTION

Does the Sun have a chromosphere in magnetic field free regions? We certainly observe enhanced emission over radiative equilibrium values, but does this mean the gas kinetic temperature increases outward above a temperature minimum? It must at some altitude since the existence of a million degree corona has long been known from the identification of coronal emission lines as arising from highly ionized atoms. Initially, the existence of a chromosphere as the onset of this rise was inferred from the sharp increase in the emission scale height of lines and continua just above the limb of the Sun. Later on, starting with the analysis of the 1952 eclipse data by Thomas & Athay (1961) and others, the chromosphere was inferred from the increase in radiation temperature with increasing opacity in lines and continua. Since then, many one-dimensional hydrostatic solar models have been constructed that reproduce the observed continuum and line intensities in an average sense (both temporally and spatially), with a chromospheric temperature rise starting at about 500 km above the photosphere. Familiar examples are the Vernazza, Avrett & Loeser (1981) models representing regions of different continuum intensity corresponding to dark cell interior to very bright network locations. However, spatially and temporally resolved observations of Ca II (e.g. Lites et al. 1993), and recently of CO (Solanki et al. 1994, Uitenbroek et al. 1994) are consistent with a large fraction of the internetwork solar surface having no temperature rise in the first 500 km above the location of the temperature minimum in the VAL models.

We present results here that show the semi-empirical chromospheric temperature rise to be an artifact of temporal averaging of the highly non-linear UV Planck function. Also, diagnosticians must be especially wary because dynamic effects can drastically alter the heights at which the dominant contributions to emission occur. We briefly describe our calculations in §2. Our results are presented in §3 and discussed in §4.

## 2. METHOD

We solve the one-dimensional equations of mass, momentum and energy conservation together with the non-LTE radiative transfer and population rate equations, implicitly on an adaptive mesh (see Carlsson & Stein 1992). The radiative transfer is treated using Scharmer's method (Scharmer 1981, Scharmer & Carlsson 1985, Carlsson 1986). The advection terms are treated using van Leer's (1977) second order upwind scheme to ensure stability and monotonicity in the presence of shocks. An adaptive mesh is used (Dorfi & Drury 1987) in order to resolve the regions (such as shock fronts) where the fluid properties are changing rapidly. The equations are solved implicitly to ensure stability in the presence of radiative energy transfer and so that the time steps are controlled by the rate of change of the variables and not by the Courant time for the smallest zones.

Our initial atmosphere is in radiative equilibrium without line blanketing above the convection zone and extends 100 km into the convection zone, with a time constant divergence of the convective energy flux (on a column mass scale). Waves are driven through the atmosphere by a piston located at the bottom of the computational domain (100 km below $\tau_{500} = 1$) whose velocity is chosen to reproduce a 3750 second sequence of Doppler shift observations in an Fe I line at $\lambda 396.68$ nm in the wing of the Ca H-line (Lites et al. 1993), formed at a height $\approx 260$ km above $\tau_{500} = 1$. There is a transmitting boundary condition at the top of the computational domain.

We include 5-level plus continuum model atoms for hydrogen and singly ionized calcium. The ionization of other species is treated in LTE, as background continua, using the Uppsala atmospheres program (Gustafsson 1973). After completing a dynamical calculation, we then re-calculate the behaviour of the C I, Si I, Mg I and Al I continua in non-LTE, assuming they do not feed back on the dynamics. These elements constitute all the relevant atoms, apart from Fe, and correspond to those considered in the VAL (Vernazza, Avrett & Loeser 1973, 1976, 1981) series of papers.

## 3. RESULTS

### 3.1. Continua Formation

To properly interpret spectra, we need to know how the observed light is produced. First, we discuss the formation of chromospheric continua. We will see that concepts that work well in a quasi-static photosphere do not necessarily work in the dynamic chromosphere.

The emergent intensity is the sum of weighted contributions involving the source function, and the column density of emitting atoms. The source function is more or less strongly coupled to the Planck function which is very temperature sensitive in the UV, although it varies only linearly with temperature in the mm wavelength range (cf Kalkofen et al. 1984).

In a dynamic atmosphere, with shocks present, the contribution function for the intensity may be doubly peaked, with one peak around $\tau_\nu = 1$ and another at a shock at smaller optical depth. In the shock the strongly enhanced source function in the UV may outweigh the small column density of emitting atoms and dominate the emergent intensity, even at very small optical depth. The mean formation height may then be somewhere between the shock and optical depth one, and have no relation to either location.

Continua formed in the photosphere, such as Al I with its edge at 207 nm, and those formed higher up till about 0.5 Mm, like Si I with its edge at 152 nm, have contribution functions peaked near $\tau_\nu = 1$ with no secondary maximum at the height where shocks exist, because the number of their atoms at that height is extremely small. The C I continuum with its edge at 110 nm is formed close to where shocks form and often exhibits a doubly-peaked contribution function (Fig. 1). In this case the exponential temperature sensitivity of the Planck function prevails, even though the number of atoms at shock forming heights is small ($\lg \tau_\nu = -4$).

Transfer of radiation decouples the source function from local conditions. As a result, the source function will typically not vary as much as the Planck function when a wave passes through (Fig. 2). We can also clearly see that the radiation temperature of the mean intensity is weighted towards the larger values of the source function and thus is a very non-linear average of the changing conditions with the passage of waves. It is important to note that there is no shock signature in the form of a saw-tooth intensity variation at any continuum wavelength. This is because the source function is non-locally dominated at the shock formation heights and because the intensity is formed over a range of heights.

### 3.2. Semi-Empirical Model Atmosphere

We analyze the dynamical simulation in a way similar to the construction of the VAL3 models. The time averaged intensity as a function of wavelength from the simulation is taken as the quantity to be reproduced by a semi-empirical, hydrostatic, model atmosphere. For an assumed temperature structure the equations of hydrostatic equilibrium, statistical equilibrium and radiative transfer are solved and the computed intensities are compared with the time average from the dynamical atmosphere. The semi-empirical temperature structure is adjusted and the process is iterated to convergence.

The same atoms used to calculate the time average of the intensity from the dynamic simulation (hydrogen, carbon, silicon, aluminium and magnesium) are treated in non-LTE. The atomic models employed are similar to those in VAL, but with updated photoionization cross-sections from Mathisen (1984) and the opacity project. All other opacity sources and electron donors are treated in LTE.

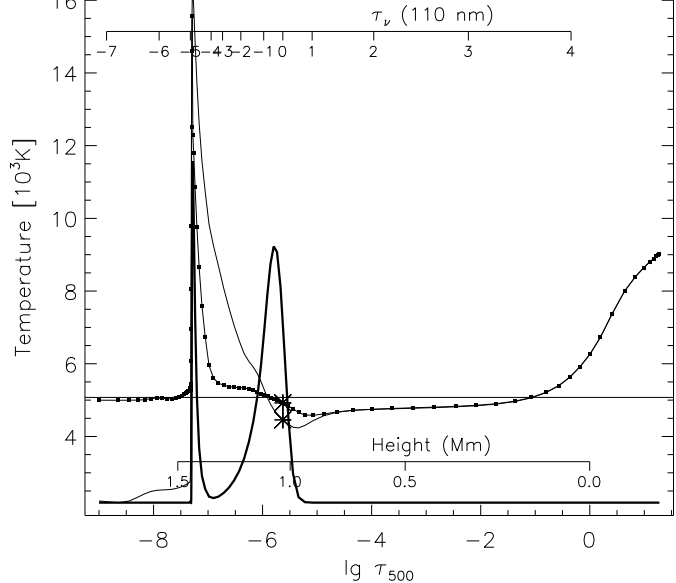

**Fig. 1.** Snapshot from the dynamical simulation. Temperature (thin solid line), source function at 110 nm (dotted) and the contribution function to the intensity (thick solid) as funtions of $\lg \tau_{500}$. The temperature and the source function at $\tau_\nu = 1$ are marked with asterisks. The radiation temperature of the outgoing intensity is shown as a horizontal line. The contribution function to the intensity has peaks both at $\tau_\nu = 1$ and in the shock front.

Shortward of infrared and mm wavelengths only UV continua are formed above the position of the temperature minimum in the VAL models. The semi-empirical temperature structure is thus determined by the time averaged UV intensities from 150 nm (formed around the VAL temperature minimum) to the Lyman continuum (formed in the upper chromosphere in the VAL models).

The final semi-empirical temperature structure is typically able to reproduce the time averaged intensity to within 20 K in radiation temperature above 100 nm and to within 100 K in the Lyman continuum below 91.2 nm.

### 3.3. Gas and Semi-Empirical Temperatures

Figure 3 shows the time average of the gas temperature as a function of height in the dynamical simulation (thick solid line) and the semi-empirical model that gives the best fit to the time average of the calculated intensities (thick dashed). Also shown are the range of temperatures in the simulation, the starting model, and the semi-empirical model FAL-A constructed to reproduce the solar dark internetwork regions (Fontenla et al. 1993).

The striking feature of Fig. 3 is that the time average of the temperature as a function of height in the dynamical simulation shows *no chromospheric temperature rise*. Instantaneously, the viscous dissipation in the shocks together with the pres-sure work leads to a high temperature. This shock spike is very strong because the long time scales for hydrogen ionization/recombination initially prevents the energy from going into ionization energy (Carlsson & Stein 1992). Further back into the post-shock region hydrogen is ionized which lowers the temperature. Even further behind the shock, recombination occurs releasing the energy in the form of radiative cooling. Integrated over time at a given height, the viscous dissipation is balanced by the radiative cooling and there is no increase in temperature. There is, however, an increase in radiation compared with the initial radiative equilibrium atmosphere due to the radiative cooling. To reproduce this increase in the intensity, the best match semi-empirical model needs a chromospheric temperature increase.

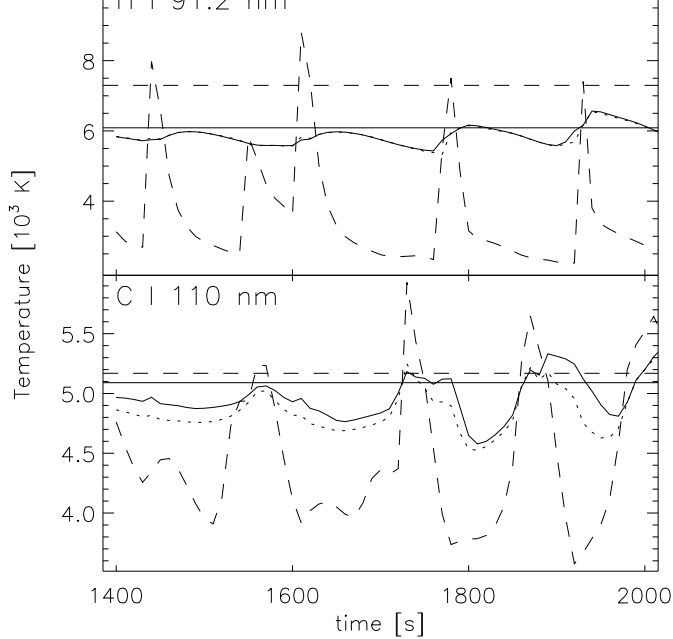
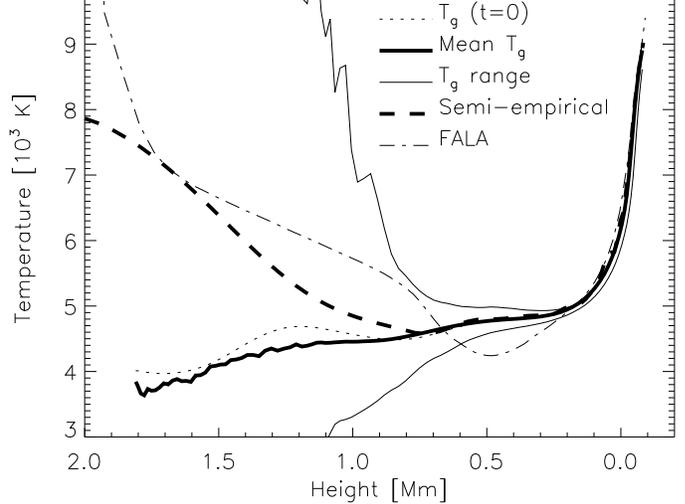

**Fig. 2.** The radiation temperature of the outgoing intensity (solid), the source function at $\tau_\nu = 1$ (dotted) and the temperature at $\tau_\nu = 1$ (dashed) as functions of time for a small part of the dynamical simulation for two wavelengths. The horizontal lines show the radiation temperature of the mean of the outgoing intensity (solid) and of the mean of the Planck function (dashed) with the mean taken over this part of the simulation. The means from the complete simulation will be slightly different. The non-linear averaging is clearly shown with the radiation temperature of the intensity mean being close to the maximum intensity and likewise for the Planck function. The source function varies much less than the Planck function due to the non-LTE decoupling. The radiation temperature of the outgoing intensity does not show the shock signature of a discontinuous rise for the same reason

**Fig. 3.** Time average of the temperature in the dynamical simulation, the range of temperatures in the simulation, the semi-empirical model that gives the best fit to the time average of the intensity as a function of wavelength calculated from the dynamical simulation, the starting model for the dynamical simulation and the semi-empirical model FALA. The maximum temperatures are only reached in narrow shock spikes of short duration. The semi-empirical model giving the same intensities as the dynamical simulation shows a chromospheric temperature rise while the mean temperature in the simulation does not.

### 3.4. Comparison with Observations

The simulations give a Ca II H-line behaviour very similar to the observed one (see Carlsson & Stein 1993, 1994). The observations clearly show the absence of a quasi-static chromospheric temperature rise; most of the time in the internetwork *there is no emission* in the line core (Lites et al. 1993). Ca II is the dominant ionization stage of calcium in a very wide temperature range under solar chromospheric conditions; in the dynamic simulations practically all calcium is in the form of Ca II all the time even though the temperature may vary from 2500K to 25 000K. Therefore, the optical depth scale in the resonance lines only depends on the column mass and is independent of temperature fluctuations. Since the source function has a large contribution from the local temperature, the absence of core emission implies there is no temperature rise up to at least a height of 1 Mm most of the time. The resonance lines of Mg II form slightly higher than those of Ca II due to the larger magnesium abundance. Spatially resolved balloon observations of these lines by Lemaire & Skumanich (1973) and Staath & Lemaire (1994) show them to be in emission over the whole disk. We ascribe this to the presence of the magnetic field which at higher altitudes can no longer be contained by the sharply decreasing plasma density.

Spatially resolved CO observations (Solanki et al. 1994 and Uitenbroek et al. 1994) are compatible with the absence of a quasi-static temperature rise in internetwork regions. The shocks in the dynamical simulation form at about 1 Mm and will probably not be visible in the CO lines that are formed deeper in. A more detailed comparison with time resolved CO observations is in preparation.

A comparison with continuum observations is not straightforward because line-blanketing is strong in the UV but is neglected in the dynamical simulations. Furthermore, there is a lack of *spatially resolved* spectra with an absolute calibration in this wavelength regime (see Brekke & Kjeldseth-Moe 1994a). The time averaged spectrum from our dynamical simulations ties. One current discrepancy is the sign and magnitude of the intensity jump over the silicon continuum edge at 152 nm where we get an intensity increase going from short to long wavelengths and observations show an intensity decrease (Brekke & Kjeldseth-Moe 1994b) most of the time. However, this feature in the computations is rather dependent on the line-blanketing (Brekke 1992).

Time resolved images at 160 nm have been obtained with the HRTS instrument (Cook et al. 1983). These observations show a small variation in the radiation temperature with a peak-to-peak variation of up to 60–80 K (Hoekzema 1994). The simulations show a rms variation of 39K, a somewhat larger variation than in the observations.

## 4. DISCUSSION

Our main result is that there is no theoretical or observational evidence for a temperature rise in the magnetic field free internetwork chromosphere. The observed enhanced emission can be produced by temporally varying waves that generate short intervals of high temperatures, without any outward increase in the average temperature. Because of the exponential dependence of the Planck function on temperature in the ultra-violet, these short intervals of high temperature dominate the time averaged intensity, even though non-LTE effects tend to reduce this sensitivity. Hence, the radiation temperature represents preponderantly the peaks in the gas temperature rather than its mean value.

The extra energy that is radiated away comes primarily from the energy dissipated by the wave motions, which goes directly into radiation without passing through a mediating state of enhanced mean thermal energy.

One should also be aware that significant differences exist between hydrostatic model atmospheres and the average state of a dynamic atmosphere. The presence of waves changes the mean state of the atmosphere, so that procedures that work well in the photosphere may fail badly in the chromosphere.

intensity contribution functions with one peak at $\tau_\nu = 1$ and another at the shock. Thus the mean height of formation for lines and continua formed around 1 Mm can vary greatly with time and does not necessarily correspond to the actual layers emitting the photons. Therefore, static formation heights and contribution functions can not be used for analyzing observations of chromospheric continua and lines from an inherently time dependent atmosphere. When waves in the chromosphere have large amplitude, linear perturbation theory is not valid since the passage of waves changes the atmosphere fundamentally.

We have compared our semi-empirical model with VAL-C and FAL-A and find similar height dependence of temperature, although our continuum radiation temperatures are too low. What about other diagnostics? One would like to compare continua and lines with sufficient opacity to place their formation around 1 Mm. Spatially and temporally resolved observations of Ca II H & K lines in internetwork regions show no emission most of the time. Hence, there can be no general chromospheric temperature rise or emission would always be present. Also, high spatial resolution observations of CO emission off the limb show no evidence for a temperature rise in the low chromosphere. What other observations might shed light on this issue? High spatial resolution interferometric observations in the mm wavelength region, where the Planck function varies linearly with temperature, should show no radiation temperature rise in magnetic field free regions.

Thus, despite long held beliefs, the Sun may not have a chromosphere in the internetwork regions, at least not one with an outward increasing temperature.

Dr. Han Uitenbroek is thanked for valuable comments on the manuscript. This work was supported by a grant from the Norwegian Research Council and by grant NAGW-1695 from the National Aeronautics and Space Administration. The computations were supported by a grant from the Norwegian Research Council, tungregneutvalget.